\documentclass[
	a4paper, %
	10pt, %
	twoside, %
]{LTJournalArticle}

\usepackage[none]{hyphenat}
\usepackage{amsmath}
\usepackage{tabularx}
\hypersetup{
    colorlinks=true,
    linkcolor=blue,
    urlcolor=blue,
    citecolor=blue
}

\runninghead{2023 Intel-HDSI DCA Collaboration: Carbon Footprint Project} %

\title{Power Consumption Patterns Using Telemetry Data} %

\author{%
	Harry Cheon\textsuperscript{1}\thanks{Corresponding author: \href{mailto:scheon@ucsd.edu}{scheon@ucsd.edu}},
        Yuyang Pang\textsuperscript{1},
        Zhiting Hu\textsuperscript{1},
        Benjamin Smarr\textsuperscript{1},\\
        Julien Sebot\textsuperscript{2},
        Bijan Arbab\textsuperscript{2},
        Ahmed Shams\textsuperscript{2}
}

\date{\footnotesize
\textsuperscript{\textbf{1}}Halicioglu Data Science Institute, University of California San Diego\\ 
\textsuperscript{\textbf{2}}Intel Corporation
}

\renewcommand{\maketitlehookd}{%
	\begin{abstract}
This paper examines the analysis of package power consumption using Intel's telemetry data. It challenges the prevailing belief that hardware choice is the primary determinant of a device's power consumption and instead emphasizes the significant role of user behavior. The paper includes two sections: Exploratory Data Analysis (EDA) and a linear model for power consumption. The EDA section provides valuable insights from Intel's telemetry data, comparing power consumption across countries, with a specific focus on power consumption patterns in the US and China. Our simple linear model affirms those patterns and highlight the possible importance of user behavior and its influence on power consumption. Ultimately, the paper underscores the need to understand power consumption patterns and identifies areas where stakeholders like Intel can make improvements to reduce environmental impact effectively and efficiently.
	\end{abstract}
}

\begin{document}

\maketitle %

\section{Introduction}

The term \emph{carbon footprint} gained popularity in the early 2000s, largely
due to British Petroleum's PR campaign and their "carbon footprint calculator"
in 2004 \cite{kaufman:cf}. This campaign shifted the blame for climate change
onto ordinary consumers while promoting the petroleum industry's agenda. Despite
the controversy, the concept of carbon footprint has gained recognition
globally, leading to efforts like the Paris Agreement in 2015, where countries
agreed to reduce carbon emissions. Governments have implemented regulations to
address the climate crisis, prompting companies to become more environmentally
conscious.

Previous research on carbon footprint lacked a clear definition. In this paper,
we adopt the definition by Wiedmann et al., which defines carbon footprint as
the total amount of carbon dioxide emissions directly and indirectly caused by
an activity or accumulated over the life stages of a product \cite{wiedmann:cf}.

Our study focuses on analyzing package power consumption using Intel's telemetry
data. Although the carbon emissions associated with 1 Wh of energy vary across
regions, changing the energy sources of a country or a state is beyond the
control of individual citizens or companies. Instead, following the principles
of reduce, reuse, recycle, the first course of action is to reduce energy
consumption. Hence understanding power consumption patterns and identifying
areas where stakeholders like Intel can improve presents itself as a pragmatic
approach.

Through this project, we have challenged the prevailing sentiment among Intel
representatives that a device's power consumption is primarily determined by the
choice of hardware. Our findings indicate that although hardware choice does
play a role, it is not the most crucial factor. Instead, user behavior emerged
as the key determinant of power consumption.

The paper is divided into two parts: Exploratory Data Analysis (EDA) and a
linear model for power consumption. In the first section, we provide insights
obtained from Intel's telemetry data. They include a comparison of power
consumption across different countries using our Median User Average Power
(MUAP) metric and, most importantly, the comparisons between power consumption
patterns in the US and China. We focused on these two countries due to their
high representation in the data and initial investigations revealing significant
variation in power consumption patterns between them. Furthermore, the US and
China are the two of the largest stakeholders in the carbon emissions
conversation - 3rd in CO2 emissions per capita\footnote{Among countries with population size
greater than 10 million} and 1st in total CO2 emissions respectively
\cite{owd:co2}.

The US and China discussion includes:
\begin{itemize}
    \item Mean user power consumption distribution
    \item Time of day power consumption comparison
    \item Day of week energy consumption comparison
    \item Power consumption by OEM comparison
    \item Power consumption distribution by turbo status
    \item Power intensity of the work comparison
\end{itemize}

The second part, which is on our linear model, elaborates on the insights from
the first section and show that user behavior is the most associated factor of
power consumption.

\section{Method}

\subsection{Dataset}

Millions of people come to the Intel website per month looking to update their
device drivers for WiFi, Bluetooth, graphics, or to update their BIOS or
Firmware. Intel Driver \& Support Assistance (IDSA) website provides an
automated software program that examines each device to discover exact versions
of CPU, motherboard, graphics cards, network interfaces, etc. as well as the
associated driver, bios, and firmware versions. IDSA then compares each device’s
discovered hardware and software version against an internal database to see if
the updated version is available and upon user demand updates the device to the
latest versions. During this process each user is asked to explicitly Opt-In or
Opt-Out of sending device usage telemetry data back to Intel with a strict EULA
limiting the use of the data to improving performance, and design.

Approximately 40\% of people Opt-In, while 60\% of people Opt-Out of sending
telemetry data back to Intel. The data from these millions of worldwide devices
is the primary source of data used in this study. Over 1,400 device specific
parameters are collected from each device (CPU power used, temperature, network
traffic, Hard Drive Disk (HDD) read/write traffic, applications launch times and
duration used, frame per second experiences for each window, hard and soft page
faults, etc. There is no personal information of any kind collected from each
device, e.g. device serial number, IP or MAC address, specific URLs visited,
email, or anything that can be used to track device specific data back to
individual users. The main purpose of this telemetry effort is to study and
understand user experiences from millions of devices as used on a daily basis.

The concept of Global Unique Identifier (GUID) is crucial in understanding the
telemetry dataset. It allows device usage data over time to be associated with
the same device without compromising the user's identity. As mentioned earlier,
Intel does not collect device IP, MAC, address, serial numbers, or any other
information, which together with other data sources may lead to forms of user
Personally Identifiable Information (PII). For this reason, we create a separate
GUID which can be used to stitch data points gathered over time to the same
device. GUIDs allow us to observe the same device behavior over time to perform
various analytics, yet preserve user level PII integrity.

On each device, the Energy Server (ESRV) collects data at regular intervals. The
collected data is then analyzed and aggregated on the client device every 24
hours. When the machine is active and connected to the network, the compressed
data is uploaded to the cloud.

\begin{figure*}[ht]
        \centering
	\includegraphics[width=\linewidth]{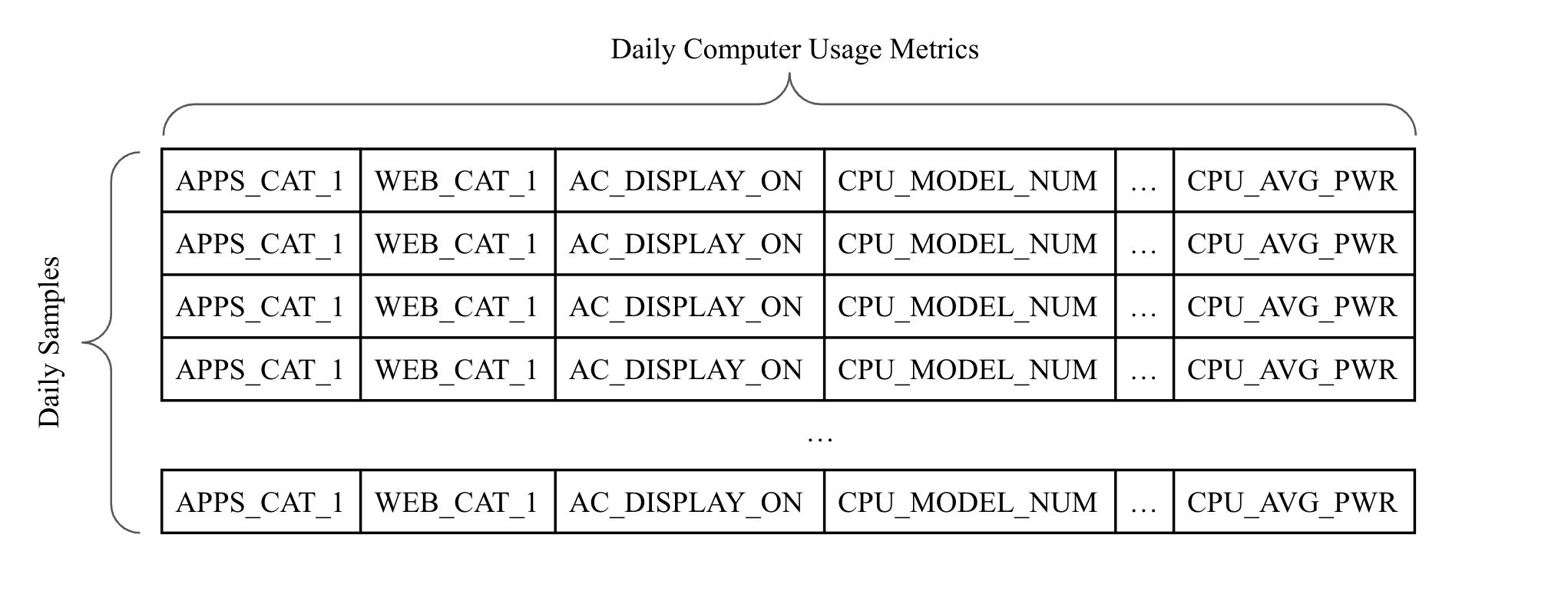}
	\caption{Illustration of the feature matrix for the linear model. Each entry
      (row) is a daily record from a machine (GUID day).}
	\label{fig:lm}
\end{figure*}

The dataset used for this study was a 1 million GUID sample of Intel's telemetry
data. Below are tables the study used for Exploratory Data Analysis (EDA):
\begin{itemize}
    \item \textbf{hw\_pack\_run\_avg\_power}: Average package power consumption for collection period
    \item \textbf{os\_c\_state}: CPU Consumption
    \item \textbf{sampler\_data}: CPU turbo data
    \item \textbf{sysinfo}: System meta information such as country, type of
        machine (laptop or desktop), type of CPU, Original Equipment Manufacturer (OEM)
        and more
\end{itemize}

Note that the data used in this study are mostly daily aggregates. This is
because data was not consistently available for narrower time intervals.

Please see the project \href{https://github.com/harrycheon/intel-cf-public}{repository} 
for a sample toy dataset.

\subsection{Data Analysis}

\subsubsection{Normalisation}

Some of the tables used in this study sometimes have multiple entries for the same
day (although the collection and upload cycle is 24 hours, it doesn't cut off at
midnight). Here, we take the weighted mean (of the metric of interest like mean package power consumption)
between same day entries, weighted by the number of samples. The weighted
mean of the metric $\omega_p$ is calculated as follows:
\begin{equation}\label{eq:wt_mean}
    \omega_p = \frac{\sum_{i=1}^{N} n_i\mu_i}{\sum_{i=1}^{N} n_i}
\end{equation}
where $N$ is the number of entries within the day, $n_i$ is the number of 
samples and $\mu_i$ is the mean metric for the $i$th entry.

We use the weighted mean for the mean CPU package power consumption and CPU
utilization aggregation.

In addition, we normalize CPU utilization by the number of cores in the system since 25\% utilization in a
4-core system is not the same as 25\% utilization in a 2-core system. To do so,
we multiplied the weighted mean daily CPU consumption by the number of cores.
Hence the normalized CPU usage $\eta$,
\begin{equation}\label{eq:norm_cpu}
    \eta = c\omega_u
\end{equation}
where $c$ is the number of CPU cores and $\omega_u$ the weighted mean CPU
utilization calculated using equation \eqref{eq:wt_mean}.

One can interpret this new feature as the workload or CPU consumption if the
machine was a single core system.

\subsubsection{Country Comparison}
We use the \emph{Median User Average Power} (MUAP) metric to compare the average
power consumption of a ``typical'' user in a country.

It is calculated as follows:
\begin{enumerate}
    \item Calculate the arithmetic mean of the average daily power consumption
        (calculated using \eqref{eq:wt_mean}) for each user
    \item Aggregate by country
    \item Take the median for each country
\end{enumerate}

\subsubsection{Turbo} \label{sec:turbo}
Intel's Turbo Boost Technology (“Turbo”) is a feature found in some Intel
processors that dynamically increases the CPU's clock speed beyond the base
frequency to provide additional performance when needed.  Since our
investigation looks at daily average power usage for a machine, we say a machine
“turbo-ed” if they were in turbo more than 10\% of the overall usage time.

\subsubsection{Work Intensity}
While we do not have an absolute metric for work intensity, we compare the
intensity of the work done by different groups by controlling for hardware and
workload (CPU consumption). We do this by binning the normalized CPU usage
and plotting the distributions for the US and China for each bin.

Note that for the turbo and work intensity investigation, we have controlled for
hardware by looking at machines with Intel\textregistered
Core\texttrademark U-series processor, which mostly have 15W Thermal Design Power (TDP).

\begin{figure*}[ht]
        \centering
	\includegraphics[width=\linewidth]{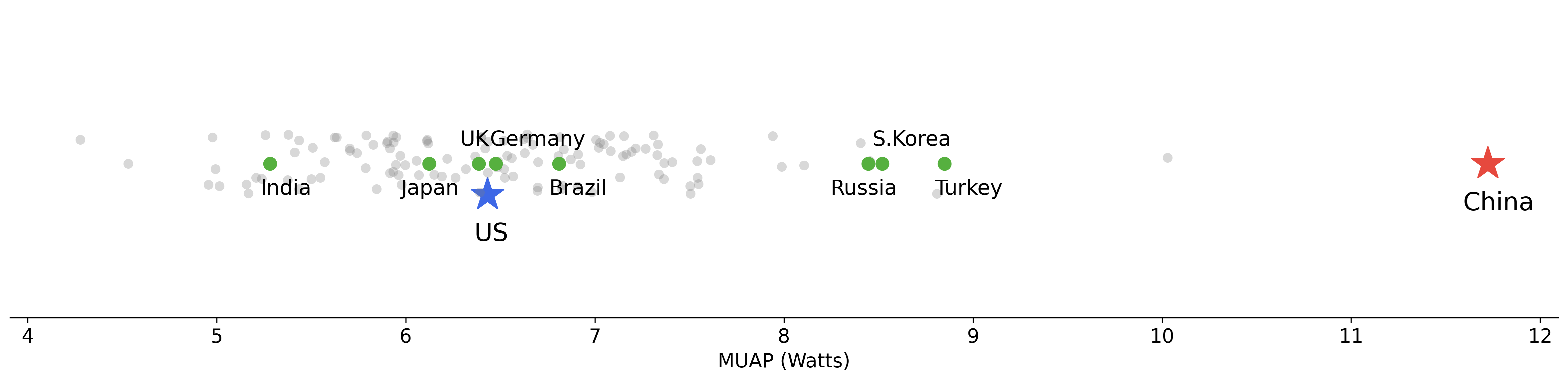}
	\caption{Median User Avaerge Power (MUAP) for countries within the dataset.
      Each scatter point is a country and the top 10 most represented countries are
      shown, with the US and China starred.  China's MUAP is almost double that of the
      US.}
	\label{fig:muap}
\end{figure*}

\subsubsection{Linear Model}

We selected a set of 225 parameters spanning 5 tables (see Appendix \ref{ap:lm-tables}) in the
aforementioned telemetry dataset. These datasets provide insights into both user
behaviors and hardware specifications such as:
\begin{itemize}
    \item Software Usage
    \item Website Usage
    \item C State
    \item ACDC Information 
    \item System Information
    \item CPU Package Power
\end{itemize}

Similar to our EDA process, the training data entries are daily aggregates (if
aggregation is applicable). As illustrated in Figure \ref{fig:lm}, each row
denotes a daily record from a distinct machine (which we call \emph{GUID day}),
while each column details the computer's usage patterns and specifications for
that day.

To obtain an overview of the data and its representation of global users, we
sampled 10,000 GUIDs (machines) selected at random. This sample served as the
foundation for building the dataset for our machine learning model, which was
constructed by querying relevant tables. Additionally, we performed data
aggregation, cleaning, and standardization. From the initial 10,000 GUIDs, we
extracted approximately 1 million daily samples.

Given our aim for the machine learning model to be interpretable, we initially
gravitated towards linear models. We primarily trained two types: LASSO and
linear regression. The LASSO model aids in identifying key features, while the
linear regression model yields precise coefficients for each feature. We also
employed other statistical measures, such as R$^2$ (coefficient of
determination), to delve into the relationship between computer usage and CPU
power consumption. To mitigate the risk of overfitting, we allocated 80\% of the
data to the training set and reserved the remaining 20\% for the test set. The
model's performance was evaluated by computing the Mean Squared Error (MSE)
based on predictions for the test set.

Please see the project \href{https://github.com/harrycheon/intel-cf-public}{repository} 
for the code ran for EDA and modelling.

\section{Results}

In Figure \ref{fig:muap}, we compare the MUAP between different countries in the
DCA database. China’s MUAP is 11.74 whereas the MUAP for the US is 6.4. In fact,
China’s MUAP is far ahead of other countries. We need to keep in mind that this
is an aggregate that does not take any confounding factors such as type of
machines (laptop or desktop) or the type of CPU. The plot below highlights the
top 10 most represented countries (including the US and China) in the dataset.

\begin{figure}[ht]
	\includegraphics[width=\linewidth]{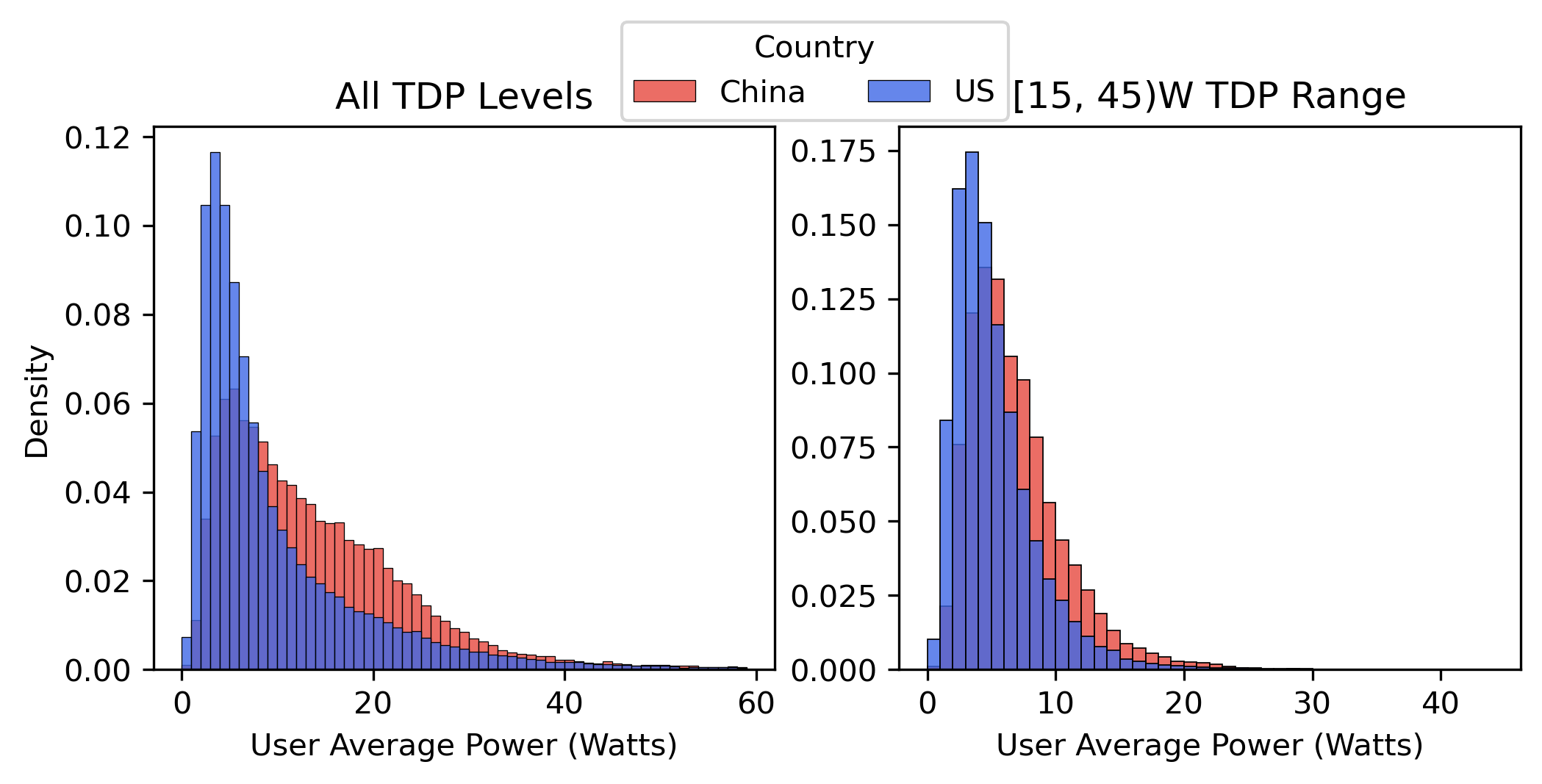}
	\caption{Histogram of User Average Power (UAP) for the US (blue) and China (red)
      with users in all TDP levels (left) and users between [15, 45)W (right).
      China's distribution of UAP is shifted to the right, even after controlling for
    TDP. Note that each data point in this plot is an aggregate for a single user.}
	\label{fig:hist}
\end{figure}
If we take a look at the US and China and their user average power (UAP)
distributions, we observe a sizable difference between the two (Figure
\ref{fig:hist}). Although the difference narrows when we control for the TDP
of the CPU, the divergence persists.

\begin{figure}[h]
	\includegraphics[width=\linewidth]{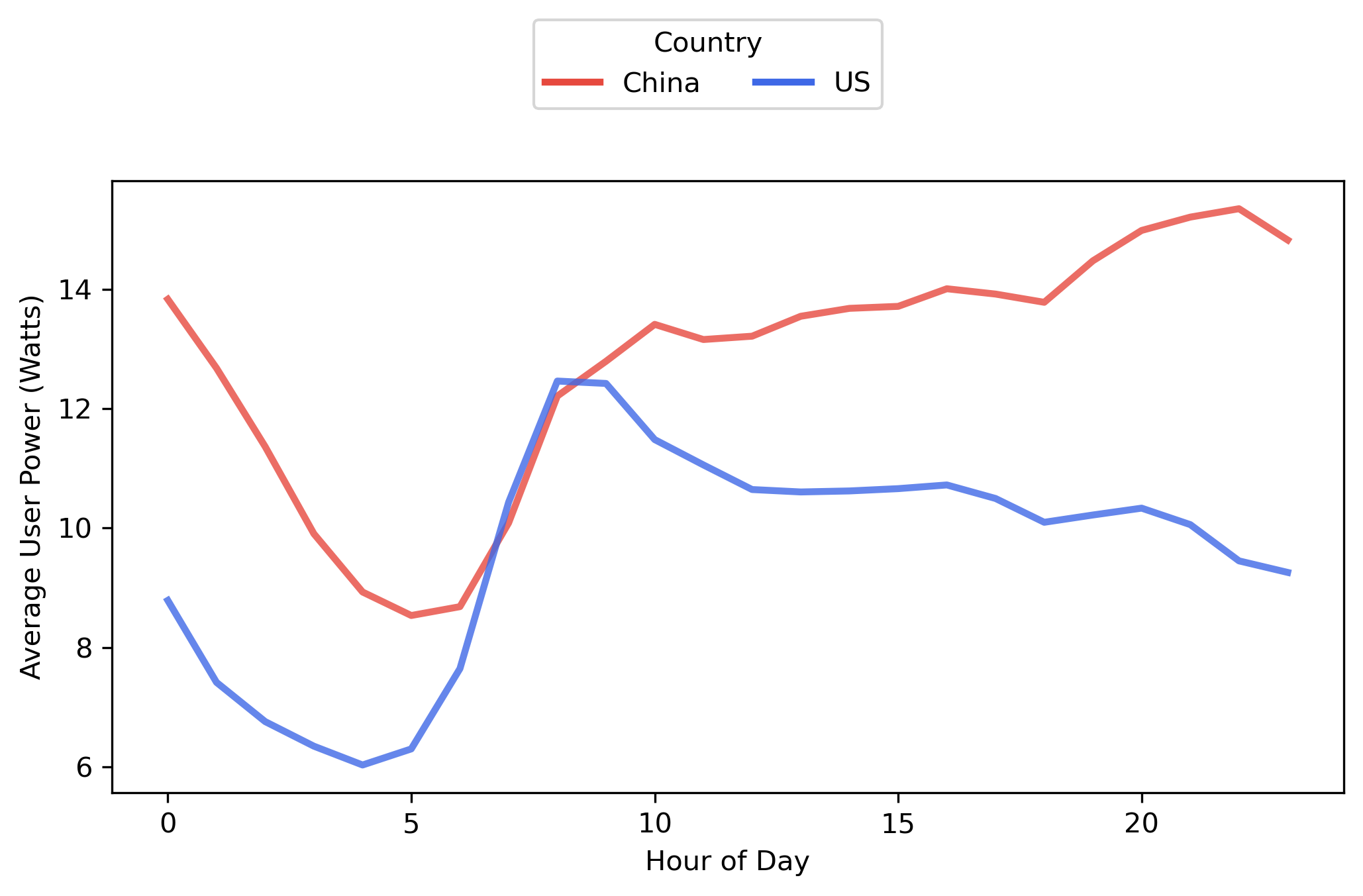}
	\caption{Average power consumption at each hour of the day for the US (blue)
      and China (red). China's power consumption increases as the day goes by,
      towards the evening, US's power consumption decreases throughout the day after
    peaking at around 8 am.}
	\label{fig:time}
\end{figure}
Figure \ref{fig:time} illustrates intraday power consumption trends for China
and the US. In China, average power consumption increases as the day goes by,
peaking at night, whereas in the US average power consumption peaks in the
morning and decreases by the hour.

\begin{figure}[h]
	\includegraphics[width=\linewidth]{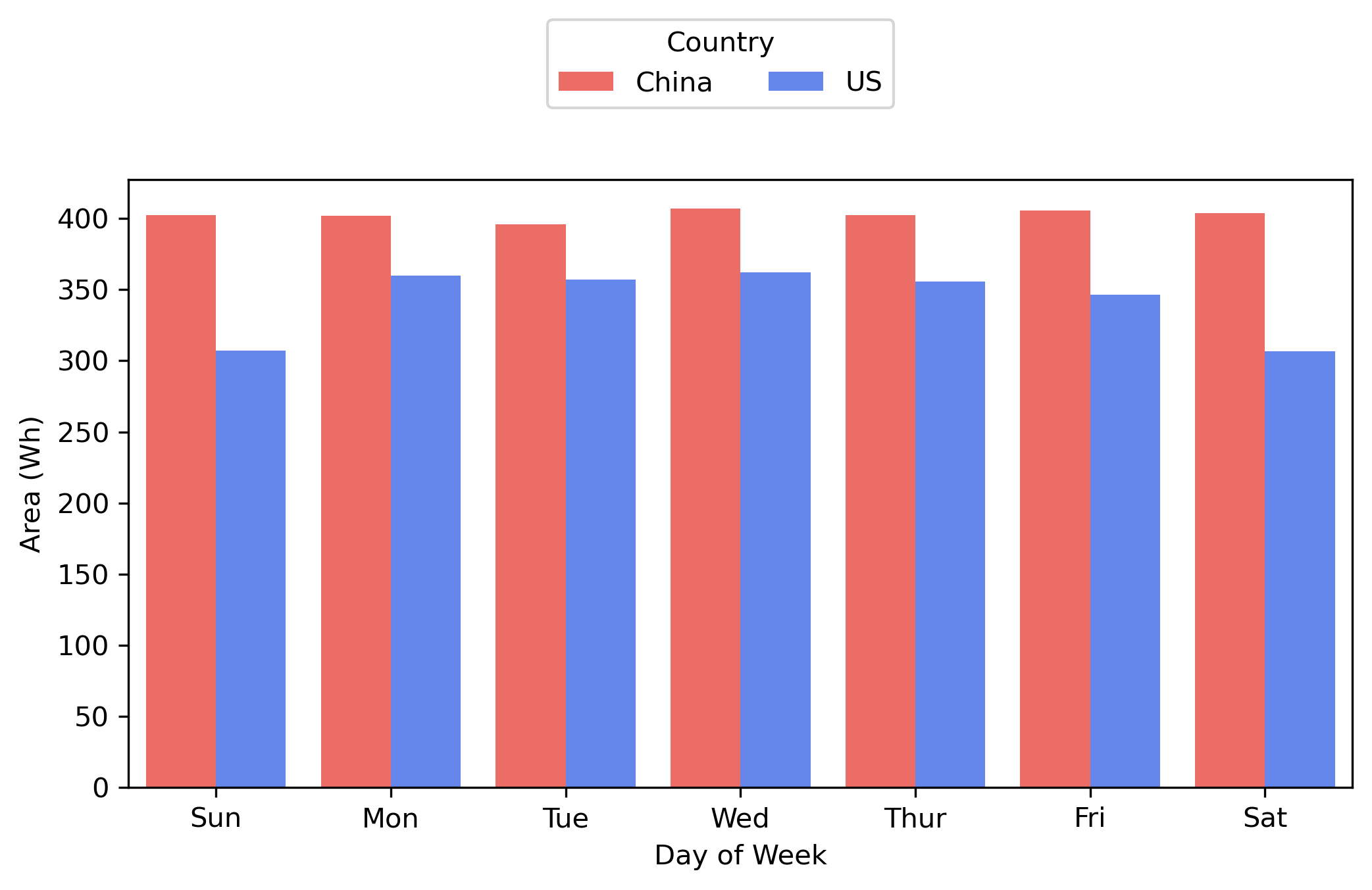}
	\caption{Area under the curve of time of day plot (like Figure \ref{fig:time})
      for each day of the week. The area under the curve represents an average user's energy consumption (Wh) on each day of the week for the US (blue) and China (red). Notice the US's area is smaller on Saturdays and Sundays, whereas China
      does not have such weekend patterns and the area stays consistent throughout all days.}
	\label{fig:dow}
\end{figure}
Figure \ref{fig:dow} shows the area under the curve of the time of day plot like
above for each day of the week. The US has a smaller area on the
weekend (Saturday and Sunday), but China has a consistent area throughout the
week.

\begin{figure}[h]
	\includegraphics[width=\linewidth]{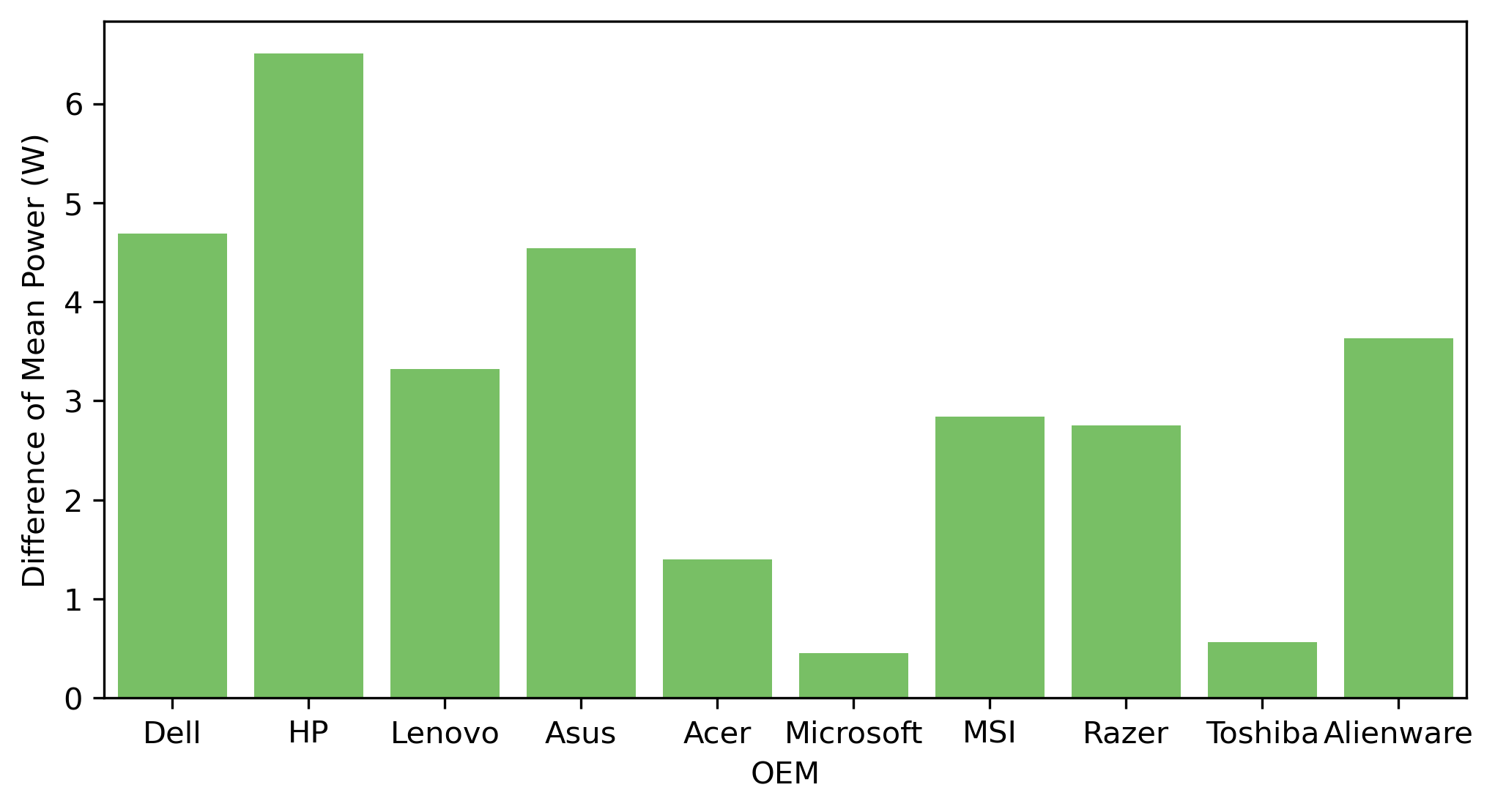}
	\caption{Difference in average power consumption between China and the US
      for the top 10 OEMs by GUID count in the US (listed in descending order by count). A positive difference indicates higher levels of average power
      consumption for China. All OEMs have significant positive difference, except
      Microsoft and Toshiba.}
	\label{fig:oem}
\end{figure}
We compare the difference between Chinese and American MUAP for laptop
manufacturers, a positive difference means that the mean UAP for the OEM is
higher in China (Figure \ref{fig:oem}). Since we are looking at laptop
manufacturers, we restricted the dataset for this particular investigation to
laptops. The plot includes the top 10 manufacturers by GUID counts in the
US (the OEMs are listed in descending order of GUID count). The difference is
positive for all manufacturers, with HP having the largest difference of 6.47
watts. We see significant differences in even gaming oriented OEMs like MSI,
Razer and Alienware.

\begin{figure}[h]
	\includegraphics[width=\linewidth]{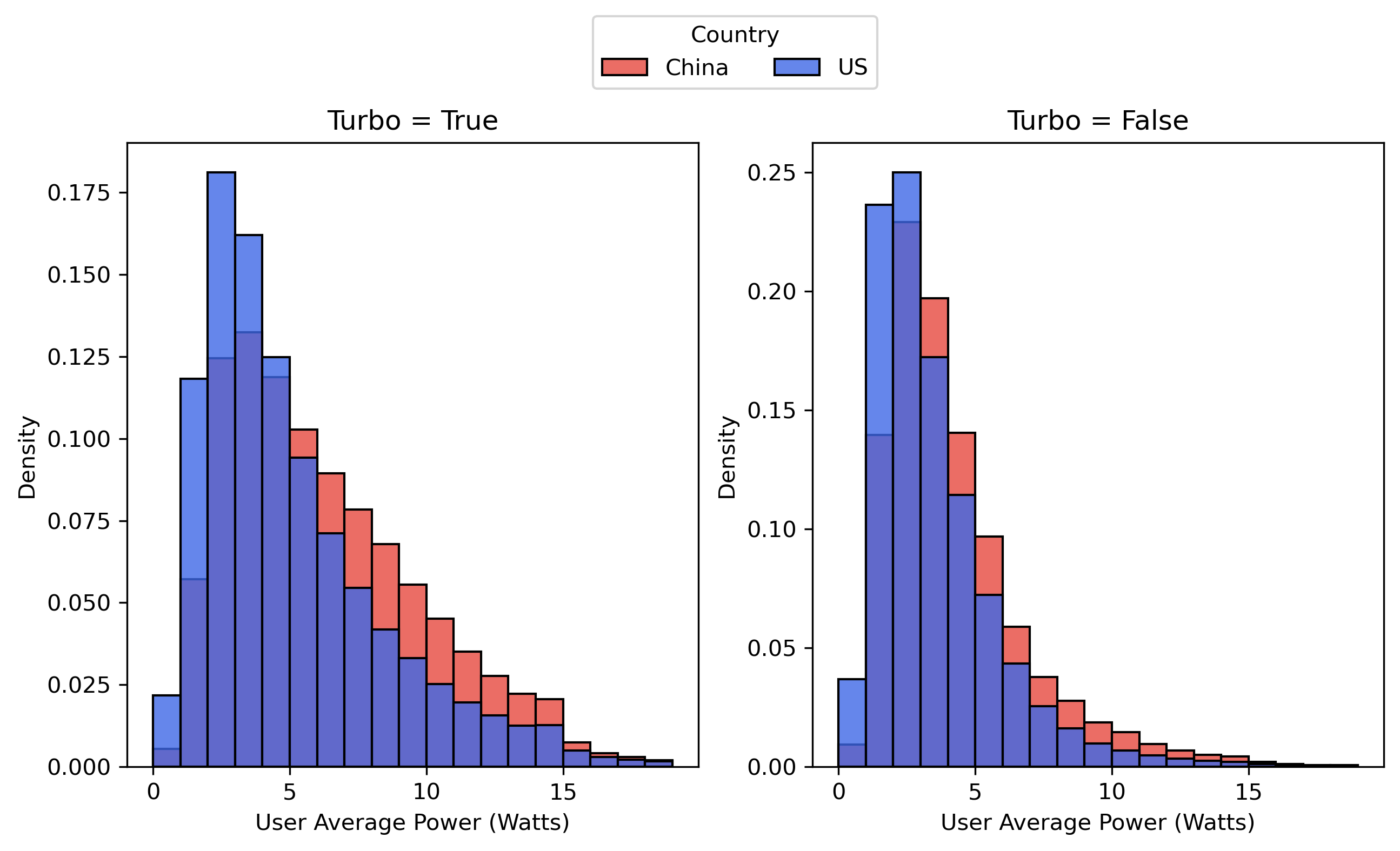}
	\caption{Distribution of daily user average power for the US (blue) and
      China (red) depending on Turbo status (see subsection Turbo on page
      \pageref{sec:turbo} for detailed definition). Difference in distribution
      between the US and China are larger when users were in turbo for that day.
      Note that each data point is a GUID day (see Subsection Linear Model
      on page \pageref{sec:turbo} for definition).}
	\label{fig:turbo}
\end{figure}
Figure \ref{fig:turbo} has the power usage distributions depending on turbo
status. Each entry in this data is a user day (or GUID day). We can see that the
difference between the two is larger when comparing turbo data for the two
countries.

\begin{figure*}[ht]
	\includegraphics[width=\linewidth]{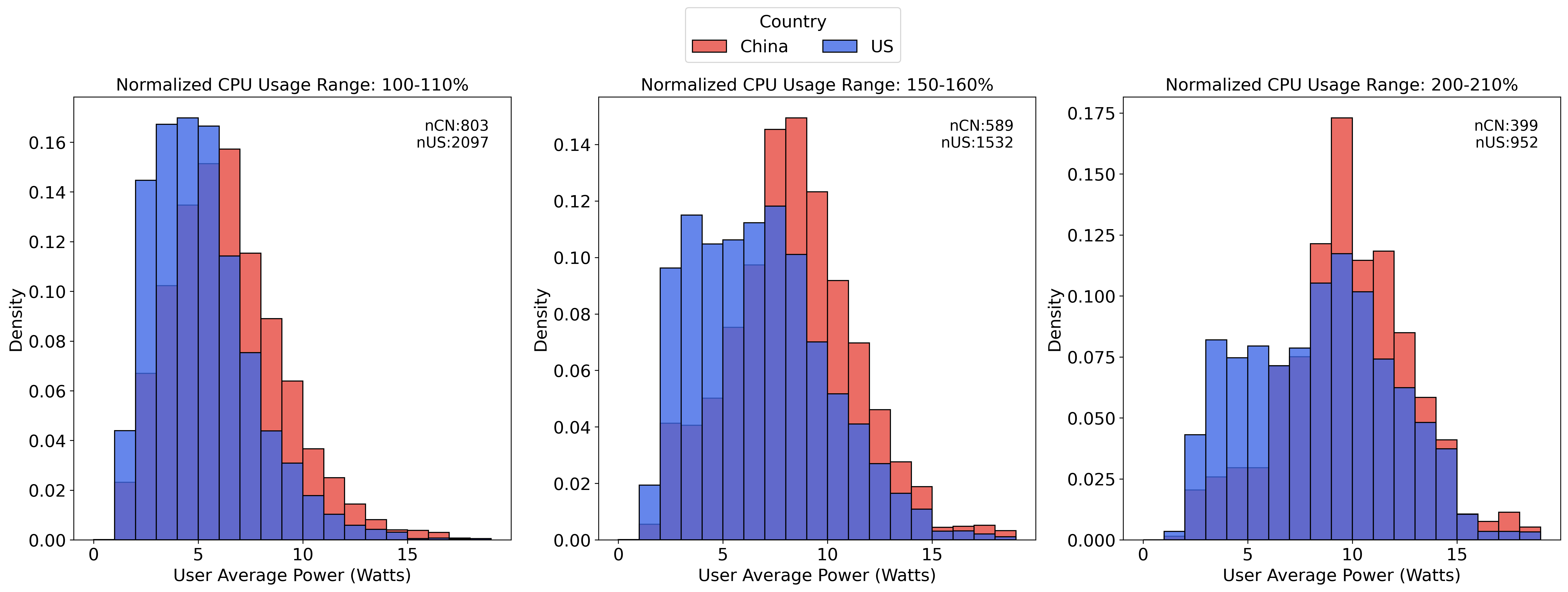}
	\caption{Histogram of daily user average power for the US (blue) and
      China (red) for each normalized CPU usage (U-series CPU machines only).
      China's distribution has a significantly smaller right-skew (or no skew) compared to the
      US in each normalized CPU Usage bin. nUS and nCN indicate the GUID counts
      in the normalized CPU usage bin for the US and China respectively. Note that
      each data point is a GUID day. See Appendix \ref{ap:intensity} Figure \ref{fig:intensity-full} 
      with plots for all CPU Usage bins.}
	\label{fig:intensity}
\end{figure*}
Figure \ref{fig:intensity} has the power consumption distributions for US and
China between the US and China at specific CPU usage intervals. We can see that
the Chinese users are consuming more power at the same CPU usage indicating
more power intensive usage.

\subsubsection{Linear Model}
The LASSO model yields a Mean Squared Error (MSE) of 5.86 watts$^2$ on the test
set. The coefficient of determination (R$^2$) for the linear regression model is
0.33. R$^2$ quantifies the proportion of variance in the dependent variable that
is predictable from the independent variable(s). Given that many factors
potentially influencing the CPU package power aren't captured in the telemetry
data or weren't included in our dataset, an R$^2$ value of 0.33 is notably
significant.

\section{Discussion}

After looking at global patterns (Figure \ref{fig:muap}), we opted to focus
our analysis to the US and China for three main reasons:
\begin{enumerate}
  \item US (13\%) and China (7\%) are the two most represented countries in
        the sample
  \item The difference in MUAP was significantly large (especially considering
        the sample size)
  \item US and China are one of the most heaviest carbon emitters and findings
        would potentially have the highest impact
\end{enumerate}

Our findings indicate that an average user from China consume twice as much
energy as their American (and global) counterparts. Even after adjusting for the
CPU Thermal Design Power, Chinese users consistently exhibit significantly higher
power consumption than those in the US. Intel's internal team confirmed that
this difference between the US and China is not limited to our 1 million GUID
sample but can also be observed in Intel's full-scale telemetry dataset.

The first thing that came to our mind were types of users in each country. Our
initial hypothesis was that there are more power intensive users in China (i.e.
gamers) than in the US. However, the difference in gamer proportion was not
significantly different (11.44\% and 10.54\% for the US and China respectively.
See Appendix \ref{ap:persona} for detailed breakdown of user personas for the two countries).

It is worth noting that there is a difference in the hardware between the two
countries. The US has more thin and light devices with CPU TDP between
$[15,45)W$ whereas China has a much larger proportion of their machines with CPU
TDP between $[45,65)W$. When we control for devices, the difference between the
two distributions narrows down, but doesn't go away entirely (Figure
\ref{fig:hist}). Therefore, the divergence between the US and China cannot be
solely attributed to types of users nor their choices of hardware.

The investigation discovered a variety of power consumption patterns between the
two countries. First we looked at time series patterns in power consumption. In
Figure \ref{fig:time}, China's average power increases towards the evening, whereas
the US's decreases over the course of the day. In figure \ref{fig:dow}, we observe
a clear weekend dip in the US, but not for China.

Since Chinese users are using more power for longer periods of time (compared to
the US) optimizing Chinese user workflows will significantly decrease total
energy use and hence have a reduction in carbon emissions.

One of our Intel investigators raised the question whether we see the divergence
between the US and China across OEMs. Most OEMs showed a positive difference in
average power between China and the US (i.e.
$\mu_{\text{CN}} > \mu_{\text{US}}$), though to varying extents (Figure
\ref{fig:oem}). As a matter of fact, HP has previously reached out to Intel
pertaining to overheating in Chinese machines.  This doesn't come as a surprise,
as HP has the most pronounced difference among all OEMs in average CPU power
consumption between the US and China.

As noted in the results section, we were surpsied by the significant difference
for gaming oriented OEMs. We expected to see a smaller difference because we
believed that gaming OEM customers are a narrow demographic that has a smaller
variance in usage patterns across regions.

The turbo investigation results also ran against our expectations. We expected
to see little difference between the two distributions when the machine was in
turbo for that day. This is because when in turbo, power consumption is limited
by hardware. As seen in Figure \ref{fig:turbo}, the difference in distribution
between the US and China was larger when the machines were in turbo. One does
need to keep in mind that each data point in the study are daily aggregates and
our threshold of whether a user was in turbo for that day is 10\%.  Although it
is quite difficult to show with the current telemetry data, we suspect that the
remaining non-turbo usage (i.e. if a user was in turbo 15\% of the day, then the
remaining 85\%) is driving the difference. Nevertheless, the results suggest
that the majority of the difference in distribution between the US and China is
driven by users that turbo more often.

The results above led us to believe that Chinese users were running more power
intensive processes. Even when we control for workload (CPU usage), we were able
to see that Chinese users were consuming more power (Figure \ref{fig:intensity}).

Consequently, we hypothesized that there might be certain usage patterns that
are unique to Chinese users that is driving the observed difference. In other
words, user behaviors compared to hardware choice may have a greater impact on
power consumption than initially anticipated. To validate this hypothesis, we
turned to our linear model. Features with larger absolute coefficient values
were deemed to have a greater impact on CPU package power.

We used L1-regularization for our linear regression (Least Absolute Shrinkage
and Selection Operator or LASSO), which penalizes the model based on the sum of
the absolute values of its coefficients. This penalty causes the absolute values
of the weights to shrink, with many being pushed to zero.  Consequently,
features with non-zero coefficients are considered pivotal for power prediction.

\begin{table}[h]
    \centering
    \begin{tabular}{|p{0.45\linewidth}|p{0.2\linewidth}|p{0.2\linewidth}|}
        \hline
        \textbf{Feature Name} & \textbf{Coefficient} & \textbf{Absolute Value}\\
        \hline
        1. CPU Normalized Usage (C state) & 1.24 & 1.24\\
        \hline
        2. Number of Cores & 0.58 & 0.58 \\
        \hline
        3. Persona Gamer & 0.42 & 0.42 \\
        \hline
        4. CPU Family Core i7 & 0.42 & 0.42 \\
        \hline
        5. Persona Web User & -0.22 & 0.22 \\
        \hline
        
    \end{tabular}
    \caption{Features with the largest absolute value of the coefficients}
    \label{table:coef}
\end{table}

As seen in Table \ref{table:coef}, among the top 5 features with the most
substantial absolute coefficients, three are related to user behavior. A
noteworthy observation is the relatively low energy consumption associated with
browsers. If this trend holds universally across various browser-based tasks,
Intel could advocate for web-based applications to users prioritizing reduced
energy consumption.

Web-based computing, also known as cloud computing, offers numerous advantages
concerning energy efficiency. Large-scale cloud data centers facilitate optimal
designs for airflow, cooling, and energy consumption. As a result, these centers
often surpass personal computers in energy efficiency on a per-operation basis
\cite{koomey2011}. Furthermore, many leading cloud providers have shifted
towards incorporating renewable energy sources \cite{greenpeace2017}.
Therefore, Promoting web-based applications can help Intel minimize its carbon
footprint globally.

\begin{table}[h]
    \centering
    \begin{tabular}{|p{0.45\linewidth}|p{0.45\linewidth}|}
        \hline
        \textbf{Highest CPU power (watt)} & \textbf{Highest Total Energy Consumption (watt-hour)} \\
        \hline
        1. Gaming (Casual, Online \& Offline) & 1. Gaming (Casual, Online \& Offline) \\
        \hline
        2. Other & 2. Other \\
        \hline
        3. Multimedia Editing (Audio \& Video) & 3. Development \& Programming (IDEs, Text Editors, Version Control) \\
        \hline
        4. Simulation \& Virtual Reality & 4. Multimedia Editing (Audio \& Video) \\
        \hline
        5. Development \& Programming (IDEs, Text Editors, Version Control) & 5. Simulation \& Virtual Reality \\
        \hline
    \end{tabular}
    \caption{Energy Consumption Ranking}
    \label{table:energy}
\end{table}

By ranking the coefficients associated with software features, we derived an
energy consumption table (Table \ref{table:energy}). To the left, we listed the
top five software categories with the highest coefficients from the linear
model. To the right, we displayed the overall energy consumption, calculated by
multiplying the coefficients by the total duration spent on each category.
Identifying the software categories that are the most energy-intensive enables
Intel to prioritize the optimization of applications within these categories and
design CPUs with enhanced energy efficiency specifically tailored for these
applications.

Our findings led to several internal investigations within Intel. One of the
findings from such internal investigations found examples of what and why
processes were leading to higher power consumption among Chinese users.

\begin{figure}[h]
	\includegraphics[width=\linewidth]{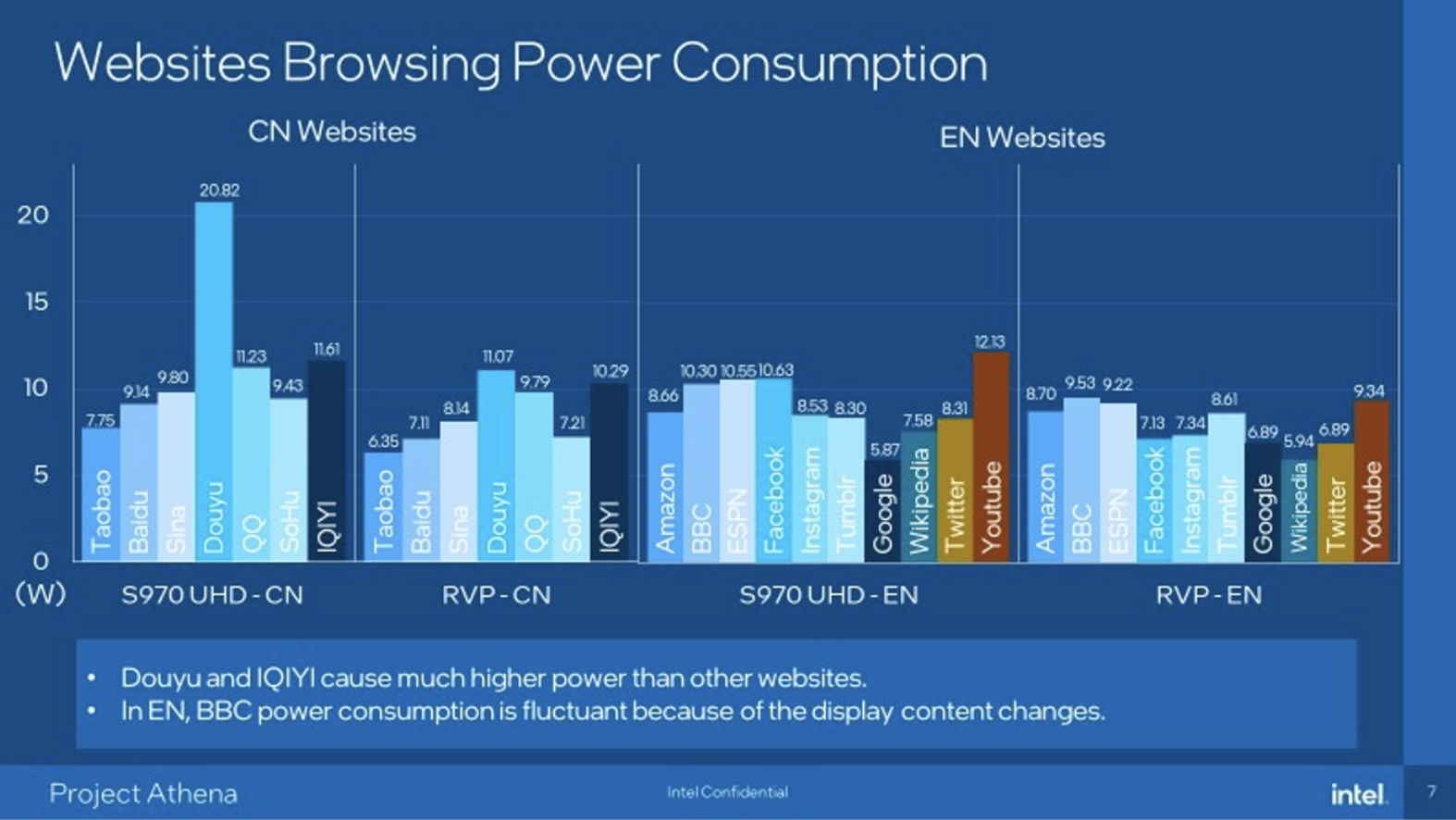}
	\caption{Intel's internal investigation results on power consumption for different
    websites. Douyu and IQIYI consumes more power than other websites.}
	\label{fig:Douyu}
\end{figure}

They discovered that DouYu, a popular Chinese streaming platform, does not
utilize Intel's Si Hardware-based DRA. Additionally, several Chinese
video/media platforms, such as Bilibili and IQIYI, don't use codec, leading them
to consume significantly more power compared to other websites running on the
same hardware (Figure \ref{fig:Douyu}), which is consistent and further validates
our findings.

\section{Conclusion}

We have shown through our analysis and modelling that user behavior has a
stronger impact on power consumption than initially anticipated. The significant
difference in MUAP between the US and China, together with the fact that they
are the two most represented countries in the dataset, presented itself as
something worth investigating. The investigation led to surprising results
pertaining to time series patterns, the effect turbo on power consumption and
power intensity of Chinese workloads.

The project aimed to contribute to Intel's 2030 sustainability goals. We were
delighted that our findings prompted internal studies at Intel and we anticipate
further initiatives to follow.

Future works include expanding the study to larger regions, incorporating carbon
emissions associated with each 1 Wh in different countries into the discussion
and developing a more complex (and accurate) model for power consumption in hopes
of creating a ``green mode'' that optimizes power consumption in real time.

\section{Limitations}
As mentioned in the methods section, unless specified (like the time of day
investigation), our work is based on daily aggregates. While this was done
out of pragmatic concerns, we acknowledge that we are losing some detail.

\section{Acknowledgement}
We would like to thank Intel and HDSI for supporting and funding this research.
In particular we would like to thank our Intel supervisors Ahmed Shams, Bijan Arbab, 
Julien Sebot and HDSI faculty supervisors Benjamin Smarr and Zhiting Hu.

\printbibliography

\newpage
\onecolumn
\appendix
\section{List of Dataset Tables Used for Linear Model} \label{ap:lm-tables}
\begin{enumerate}
    \item \textbf{hw\_pack\_run\_avg\_power}: Average package power consumption for collection period
    \item \textbf{os\_c\_state}: CPU Consumption
    \item \textbf{sysinfo}: System meta information such as country, type of
        machine (laptop or desktop), type of CPU, Original Equipment Manufacturer (OEM)
        and more
    \item \textbf{web\_cat\_usage\_v2}: Website usage information, including browser types, content categories, and duration
    \item \textbf{frgnd\_backgrnd\_apps\_v4\_hist}: Software usage information, contains software names, duration, AC/DC status, display ON/OFF, and more
\end{enumerate}

\section{User Persona Breakdown for the US and China} \label{ap:persona}

\begin{table}[h]
    \centering
        \begin{tabular}{|l|c|c|}
        \hline
        \textbf{Persona}        & \textbf{China (\%)}    & \textbf{US (\%)} \\
        \hline
        Web User                & 8.35                   & 27.48                                \\
        \hline
        Unknown                 & 34.62                  & 17.24                                \\
        \hline
        Gamer                   & 10.54                  & 11.44                                \\
        \hline
        Casual User             & 22.61                  & 10.96                                \\
        \hline
        Communication           & 2.13                   & 8.27                                 \\
        \hline
        Casual Gamer            & 3.92                   & 6.68                                 \\
        \hline
        Office/Productivity     & 6.49                   & 5.14                                 \\
        \hline
        Content Creator/IT      & 4.17                   & 5.07                                 \\
        \hline
        Win Store App User      & 1.71                   & 3.97                                 \\
        \hline
        Entertainment           & 4.38                   & 1.95                                 \\
        \hline
        File \& Network Sharer  & 1.08                   & 1.79                                 \\
        \hline
        \end{tabular}
    \caption{Persona breakdowns for the US and China. Persona was determined by Intel's Data Science team
    through k-means clustering.}
    \label{tab:persona}
\end{table}

\newpage
\section{Work Intensity Investigation Plot (Full)} \label{ap:intensity}
\begin{figure*}[h]
        \centering
	\includegraphics[scale=0.21]{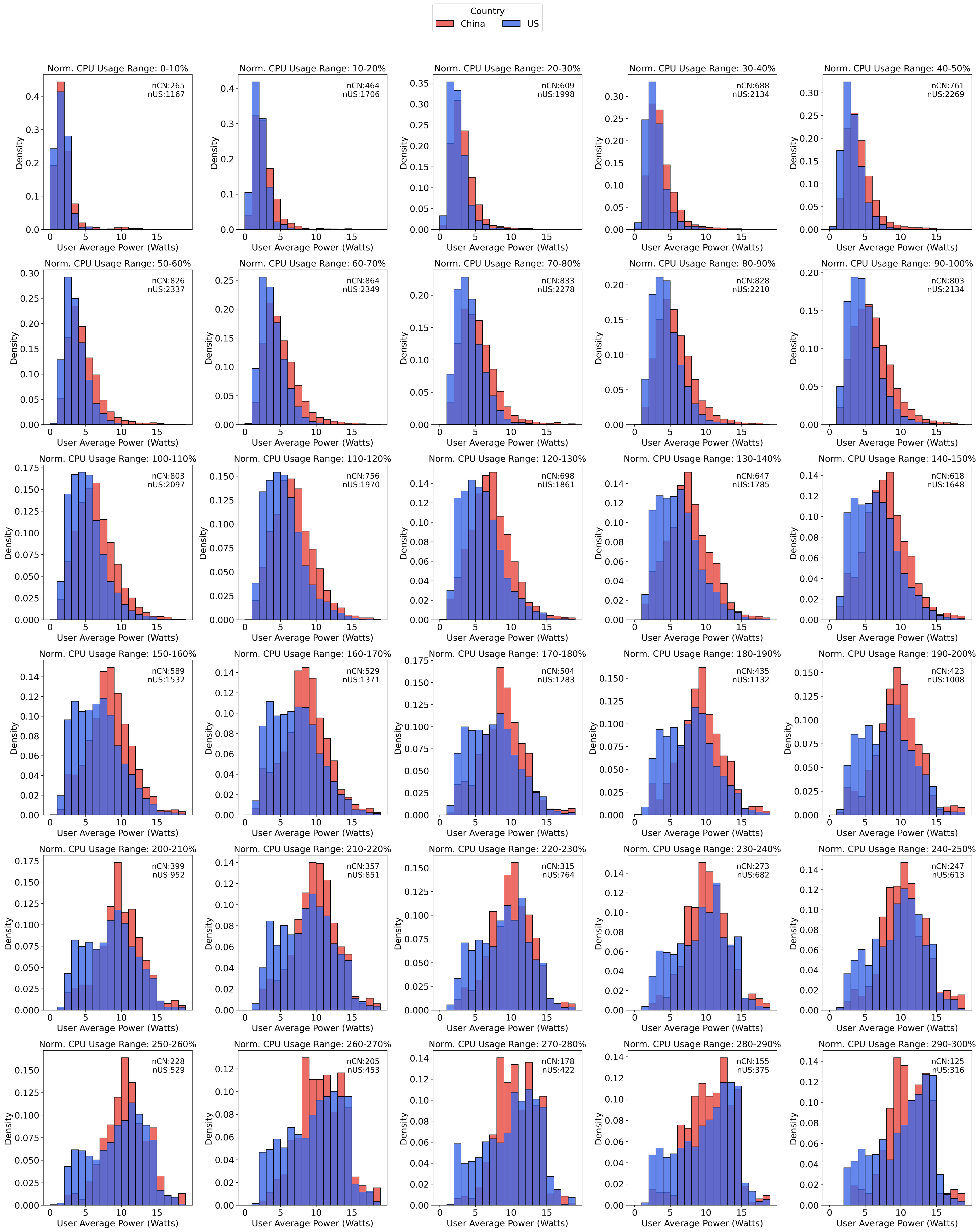}
	\caption{Histogram of daily user average power for the US (blue) and
      China (red) for each normalized CPU usage (U-series CPU machines only).
      China's distribution has a significantly smaller right-skew (or no skew) compared to the
      US in most normalized CPU Usage bins (except towards extremely high CPU usage bins). 
      The nUS and nCN indicate the GUID counts
      in the normalized CPU usage bin for the US and China respectively. Note that
      each data point is a GUID day.}
	\label{fig:intensity-full}
\end{figure*}

\end{document}